%% file: main.tex
\documentclass[sigconf]{acmart}

\usepackage{amsmath,amsfonts}
\usepackage{algorithmic}
\usepackage{graphicx}
\usepackage{textcomp}
\usepackage{xcolor}
\usepackage{url}
\usepackage{microtype}
\usepackage{booktabs}
\usepackage{float}
\usepackage{subcaption}
\usepackage{enumitem}

\input{PeterMacros}
\usepackage{tcolorbox}

\setcopyright{none} 
\acmConference[]{AI-Assisted SQL}{Meta Platforms Inc.}{2024}
\acmYear{2024}
\copyrightyear{2024}
\newlength{\imagewidth}
\input{Macros}

\begin{document}

\begin{abstract}

\sa brings generative AI into the data analytics domain. SQL is declarative, has formal table schemas, and is often written in a non-linear manner. We address each of these challenges and develop a set of models that shows the importance of each problem. 
We first develop an internal SQL benchmark to perform offline tests at \Meta. We evaluate how well the Public \llama model performs. We attain a BLEU score of 53\% and 24\% for single- and multi-line predictions, respectively. This performance is consistent with prior works on imperative languages.
We then fine-tune \llama on our internal data and database schemas. \saSchema substantially outperforms \llama by 16 percentage points on BLEU score. 
SQL is often written with multiple sub queries and in a non-sequential manner. We develop \saFIM which is aware of the context before and after the line(s) that need to be completed. This fill-in-the-middle model outperform \saFIM by 35 percentage points. We also measure how often the models get the correct table names, and \saFIM is able to do this 75\% of the time a major improvement over the other two models.
Aside from our scientific research, we also roll out \saFIM at \Meta. \sa has is used on a weekly basis by over 10k users including data scientists and software engineers, less than 1\% of users have disabled \sa. We use the feedback from users to improve \sa. Interesting positive themes include completing tedious or repetitive SQL clauses, suggesting boilerplate
coding, and help in eliminate the need to remember difficult SQL syntax. The most significant negative themes was table and column name hallucinations, which has been reduced with the release of \saFIM. 
The \sa models consistently outperform public and internal LLMs despite their smaller size (7 bn and 13 bn), which provides early indications that smaller specialist models can outperform larger general purpose models. 
\end{abstract}


\author{Chandra Maddila}
\affiliation{
  \institution{Meta Platforms Inc.}
  \country{USA}
}

\author{Negar Ghorbani}
\affiliation{
  \institution{Meta Platforms Inc.}
  \country{USA}
}

\author{Kosay Jabre}
\affiliation{
  \institution{Meta Platforms Inc.}
  \country{USA}
}

\author{Vijayaraghavan Murali}
\affiliation{
  \institution{Meta Platforms Inc.}
  \country{USA}
}

\author{Edwin Kim}
\affiliation{
  \institution{Meta Platforms Inc.}
  \country{USA}
}

\author{Parth Thakkar}
\affiliation{
  \institution{Meta Platforms Inc.}
  \country{USA}
}

\author{Nikolay Pavlovich Laptev}
\affiliation{
  \institution{Meta Platforms Inc.}
  \country{USA}
}

\author{Olivia Harman}
\affiliation{
  \institution{Meta Platforms Inc.}
  \country{USA}
}

\author{Diana Hsu}
\affiliation{
  \institution{Meta Platforms Inc.}
  \country{USA}
}

\author{Rui Abreu}
\affiliation{
  \institution{Meta Platforms Inc.}
  \country{USA}
}

\author{Peter C. Rigby} 
\authornote{Rigby is also a professor at Concordia University in Montreal, QC, Canada.}
\affiliation{%
  \institution{Meta Platforms Inc.}
  \country{USA}
}

\renewcommand{\shortauthors}{C. Maddila, N. Ghorbani, K. Jabre, E. Kim, V. Murali, O. Harman, D. Hsu, R. Abreu, P.C. Rigby}

\title{AI-Assisted SQL Authoring at Scale}
\title{AI-Assisted SQL Authoring at Industry Scale}

\maketitle

\section{Introduction}
\label{sec:intro}
While Large Language Models (LLMs) have been used extensively on general coding problems~\cite{codecompose2023,codewhisperer}, previous work on LLMs for SQL is very limited. Related work has been concerned with the generation of SQL code from natural language specifications~\cite{zhang2020m,li2024pet,zhang2024benchmarking}, but, to the best of our knowledge, no work exists on autocompleting SQL queries. 

While it can be tempting to cast SQL as just another programming language for existing LLM autocompletion systems, there are three main challenges that warrant special treatment for SQL.

\begin{itemize}[leftmargin=*]
\item[(i)] SQL is declarative in nature, and thus represents a different paradigm from the general mix of languages used in code LLM training data. SQL is intimately tied to a data warehouse (often proprietary), which severely limits the LLM's ability for knowledge transfer across training stages and programming languages.

\item[(ii)] SQL exacerbates the hallucination problem often seen in LLMs. For example, it is easy for an LLM to conjure up a non-existent table name, which then corrupts the remaining context for suggesting column names, joins, etc., quickly making the entire query invalid. While code LLMs also hallucinate, the impact of a single incorrect generation is more cascading for SQL. We conjecture that the data/db structure can help in mitigating hallucinations.

\item[(iii)] SQL queries are often written ad-hoc, from scratch, as one-off queries. There is no repository of queries as there is code, so LLMs have to work with very little context. Developers tend to write short queries often starting with the {\tt FROM} clause and working back to the final columns and aggregations. Developers will often have complex subqueries, \ie the {\tt WITH} clause, which makes the context before and after the current line even more important than in procedural programming languages. 

\end{itemize}

\begin{figure}
    \centering
    \includegraphics[width=0.4\textwidth]{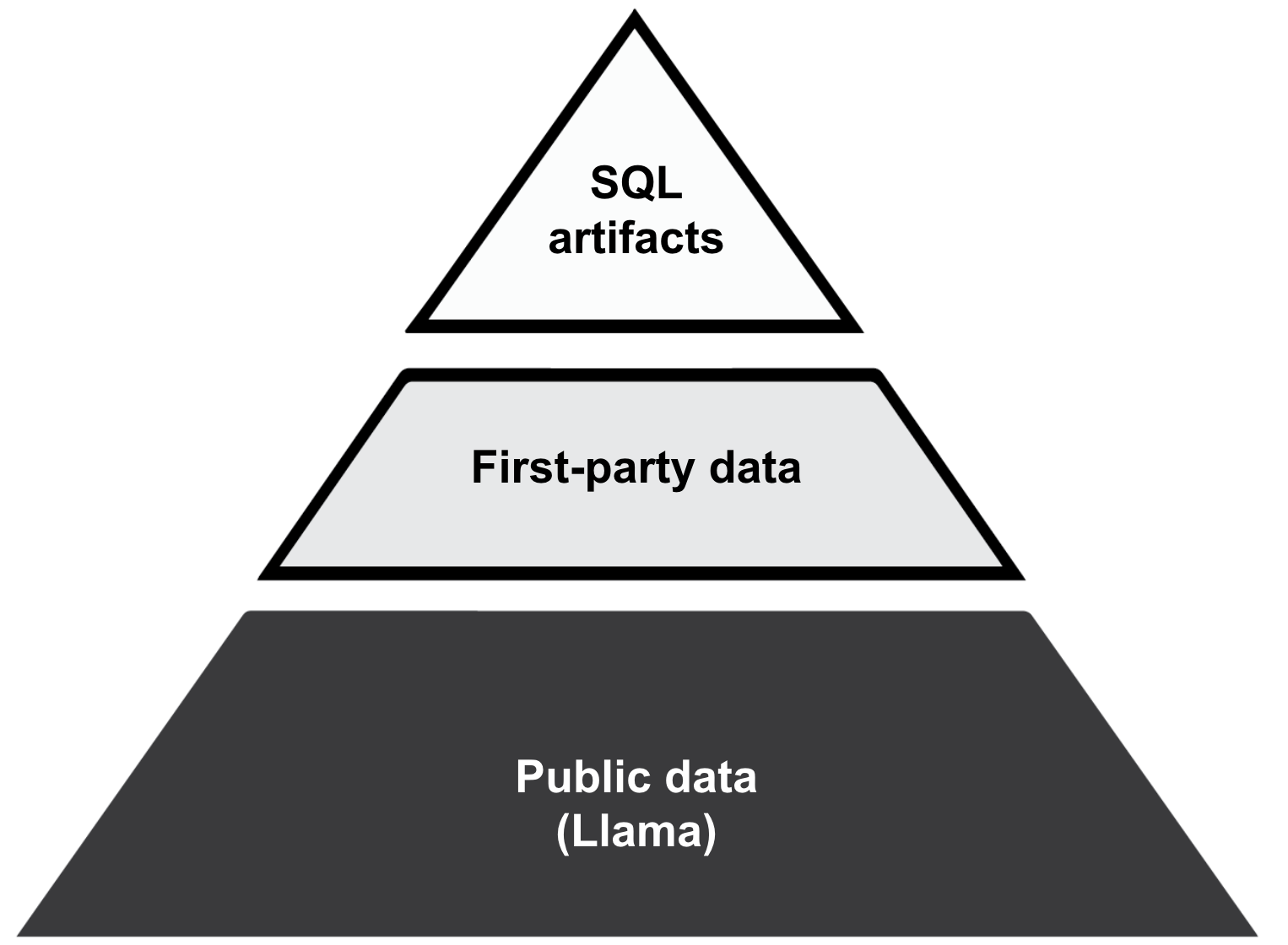}
    \caption{Data pyramid}
    \label{fig:pyramid}
\end{figure}

In this work, we provide evidence to four research questions that deal with these specific SQL challenges. Our first three research questions revolve around three candidate models and their efficacy for SQL autocompletion. For RQ 1, the pyramid in Figure~\ref{fig:pyramid} is built upon the public \llama~\cite{roziere2023code} model. For RQ 2, we then use first-party data at \Meta, \ie internal code and SQL code as well as schema information including table names and columns to create \saSchema. For RQ 3, we observe that when engineers write SQL, they often do not write in a sequential top-to-bottom manner. For example, they might fill in a {\tt WHERE} clause before fully indicating the columns to be selected. As a result, we develop a fill-in-the-middle model, \saFIM, that has context before and after the line(s) of code to be completed. Our final research question, describes the rollout to \sa to thousands of engineers and examines their feedback.

\textbf{RQ 1. Public \llama: How well does the a public model generate SQL code?}

While LLMs have been used extensively on general coding problems \cite{codecompose2023, 10.1145/3520312.3534864, li2023starcoder}, our goal is to understand how well they work on SQL, which is declarative. To establish a baseline model at \Meta, we evaluate how well the public \llama~\cite{roziere2023code} model performs on our internal benchmarks.
The \llama model is trained on publicly available code including natural language datasets related to code. \llama is also the base of our pyramid in Figure~\ref{fig:pyramid}. 

\textit{Results summary.} 
For \llama we see an exact match, BLEU, containment, and table match of 29\%, 53\%, 66\%, and 12\% for single line. The corresponding values for multi-line are 0\%, 12\%, 57\%, and 26\%, respectively. These results are comparable with prior work examining imperative languages like python~\cite{codecompose2023}. 

\textbf{RQ 2. \saSchema: How important is fine-tuning on table schemas?}

We train \saSchema on first-party data and code at \Meta. Since LLMs notoriously hallucinate, we also fine-tune on internal table schema including table names and columns. We expect to see fewer incorrect column names and invalid table names.
Figure~\ref{fig:examples} illustrates how important it is to have knowledge of the schema. The \llama model hallucinates table names and suggests infrequently used tables. In contrast, once the table names are known, the columns are correct.

\textit{Results summary.} 
For \saSchema we see an exact match, BLEU, containment, and table match of 48\%, 69\%, 78\%, 13\% for single line. The corresponding values for multi-line are 0\%, 24\%, 77\%, and 62\%, respectively. These results represent a substantial 11 to 48 percentage point improvement over the public \llama model.

\textbf{RQ 3. \saFIM: How well does a fill-in-the-middle (FIM) model perform?}
SQL authoring is usually not linear and sequential. When authoring long queries, it is common for developers to jump around nested subqueries and common table expressions (CTEs), such that information in the suffix becomes just as important as information in the prefix. See the nested query in Figure~\ref{fig:examples}.
The Fill-In-the-Middle (FIM) paradigm \cite{fried2023incoder, aghajanyan2022cm3} helps widen the context aperture. With FIM, we provide both the prefix and the suffix to the model.

\saFIM benefits from not only knowing the schema but also knowing how common the use of a particular column or table is. In Figure~\ref{fig:examples}, we see that although \saSchema has correct column names given the table, \saFIM has more context about which columns make the most amount of sense given the context of the query. 

\textit{Results summary.} 
For \saFIM we see an exact match, BLEU, containment, and table match of 50\%, 69\%, 78\%, 23\% for single line. The corresponding values for multi-line are 20\%, 59\%, 82\%, and 75\%, respectively. The improvement in single line is mostly contained to better table match percentages over \saSchema, the multi-line improvement is dramatic, increasing from 0\% exact matches to 20\%. Furthermore, \saFIM suggests the correct table 75\% of the time. 

\textbf{RQ 4. Adoption and Feedback: How is \sa used in practice?}
At \Meta, we incrementally rolled out the models. We did not rollout all models, instead, those that performed the best in the offline historical tests were rolled out.

We describe the rollout methodology and results. While we do not conduct a controlled experiment, we allow developers to provide feedback on \sa. Since there is no requirement to provide feedback, we also report the opt-out rate for \sa to ensure that developers who did not enjoy \sa but did not comment are captured. 

\textit{Results summary.} \sa has is used on a weekly basis by over 10k users including data scientists and software engineers, less than 1\% of users have disabled \sa. We use the feedback from users to improve \sa. Interesting positive themes include completing tedious or repetitive SQL clauses, suggesting boilerplate
coding, and help in eliminate the need to remember difficult SQL syntax. The most significant negative themes was table and column name hallucinations, which has been reduced with the release of \saFIM. Other negative themes include interfering with traditional auto-complete system and changes in the keyboard shortcuts and the stylistic aspects.

\section{Background and Model}
\label{sec:model}

\begin{figure}
\centering
\begin{subfigure}{1\columnwidth}
\includegraphics[width=1\columnwidth]{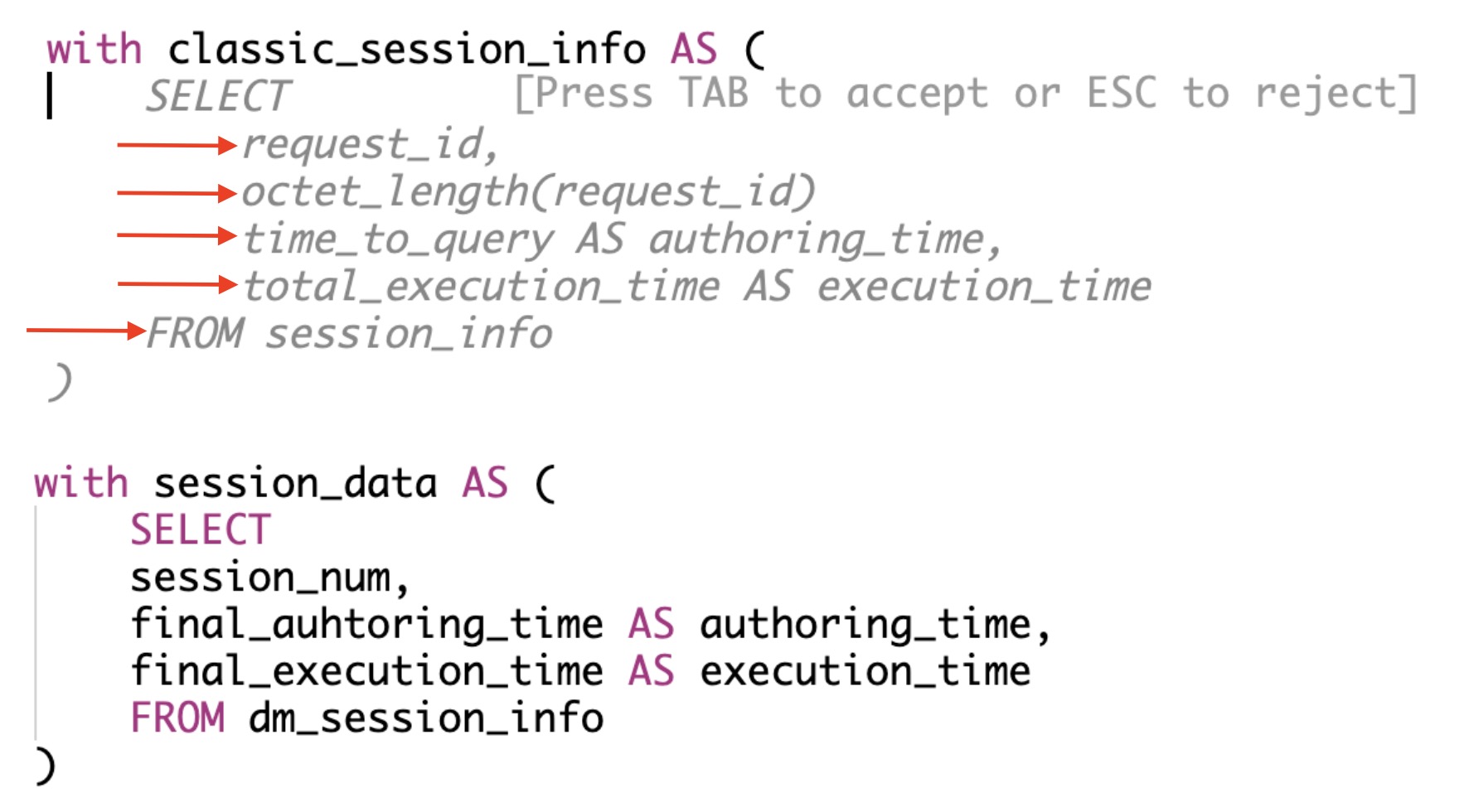}
\caption{Suggestions from public Llama model}
\end{subfigure}

\bigskip

\begin{subfigure}{1\columnwidth}
\includegraphics[width=\columnwidth]{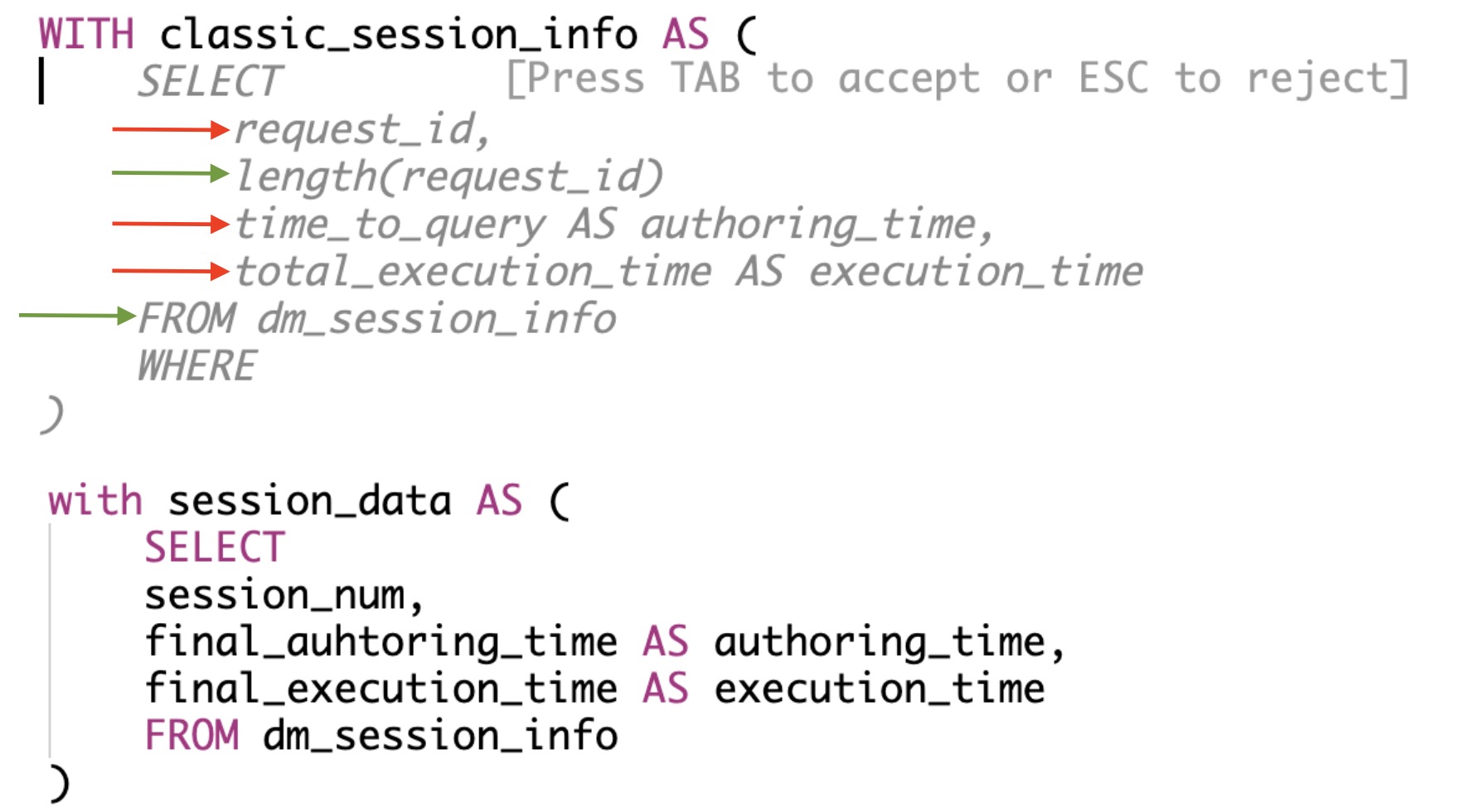}
\caption{Suggestions from \saSchema that is fine-tuned on the schema and first party data}
\end{subfigure}

\bigskip

\begin{subfigure}{1\columnwidth}
\includegraphics[width=\columnwidth]{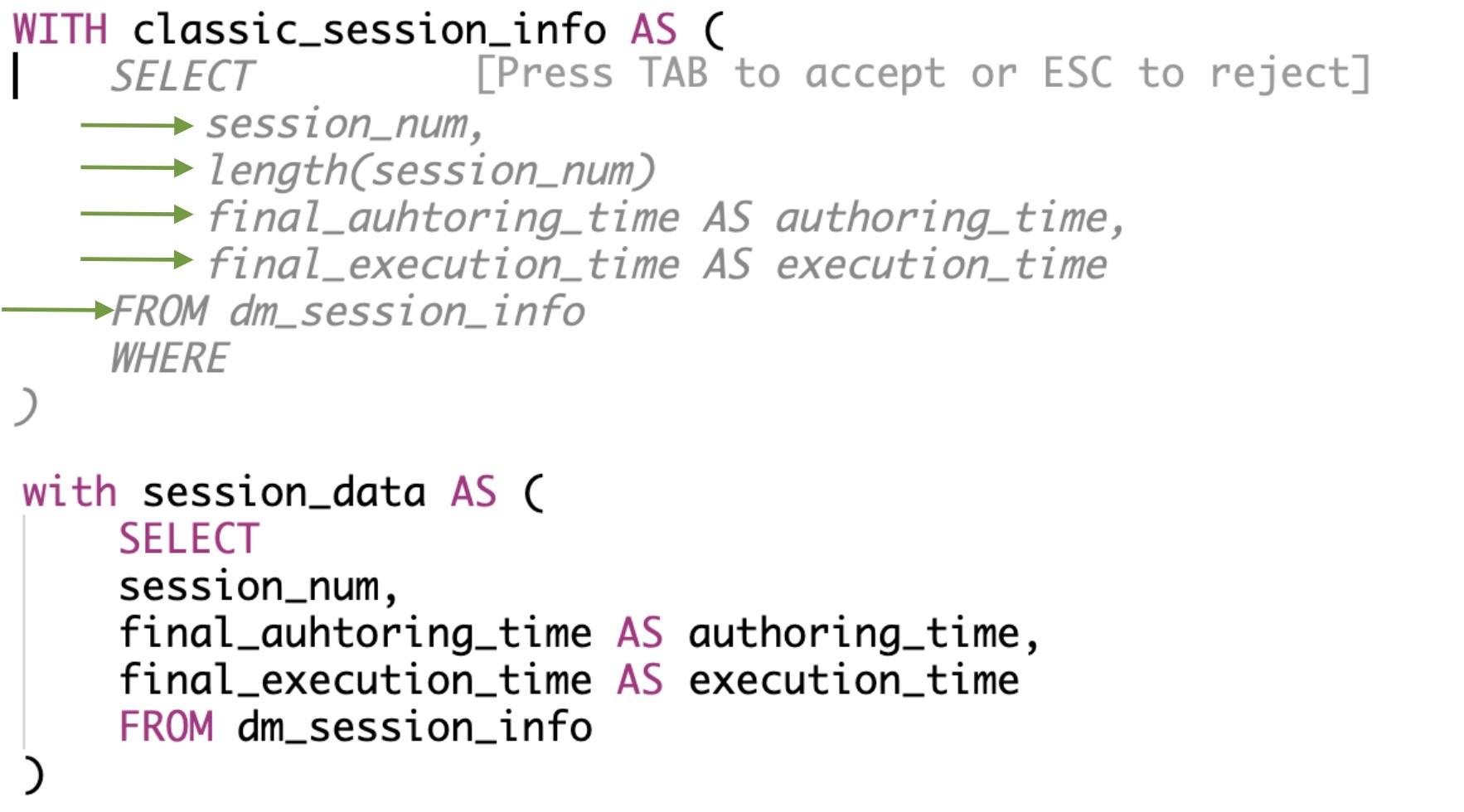}
\caption{Suggestions from \saFIM model with FIM training}
\end{subfigure}
\caption{The three screenshots demonstrate how \sa has improved its accuracy with first-party fine tuning and FIM training. The first image shows a suggestion from the public Llama model, the second image from \saSchema, and the third is generated by \saFIM with FIM training. Red arrows indicate the mistakes made by the model and green arrows indicate the correct predictions. One can see how these models  recovered gradually from their mistakes and hallucinations with training improvements (from Llama to \saSchema to \saFIM)}
\label{fig:examples}
\end{figure}

Before we provide the technical details, we introduce our running example. Figure \ref{fig:examples} shows three screenshots of a suggestion generated by plugging in three different models into the Daiquery UI  while providing the same context.

The first screenshot shows the suggestion generated by the public Llama model. It hallucinates column names, suggests a function that is nonexistent in the first-party data warehouse (\texttt{octet\textunderscore length}), and hallucinates the table name as well (\texttt{session\textunderscore info}). As a result, the generated query does not even compile and user rejects the suggestion.

The second screenshot shows a suggestion generated by the \saSchema model that is fine tuned on the first-party data warehouse. The model is already doing a good job with respect to predicting the correct function names (\texttt{length} as opposed to \texttt{octet\textunderscore length}) and table names (\texttt{dm\textunderscore session\textunderscore info} as opposed to \texttt{session\textunderscore info}). Because the model has seen numerous examples of first-party SQL and seen schema information during  training. This query compiles but generates run time errors if executed as-is because it is predicting the column names that do not exist in the table \texttt{dm\textunderscore session\textunderscore info} (\texttt{request\textunderscore id}, \texttt{time\textunderscore to\textunderscore query}, and \texttt{total\textunderscore execution\textunderscore time}). User may accept this query with an understanding to rework it to correct the column names.

The third screenshot shows a suggestion generated by the \saFIM model trained with the FIM objective. By virtue of having the bidirectional context (code before and code after), the model is able to predict the right column names as well (\texttt{session\textunderscore num}, \texttt{final\textunderscore authoring\textunderscore time}, and \texttt{final\textunderscore execution\textunderscore time}). This query compiles and executes as-is without the users needing to make any changes to the generated query.

\subsection{Data Pyramid and Models}

The data pyramid consists of four main components: A checkpoint of the public Llama model, first party data, domain specific data (SQL), and instruct fine tuning data.

\subsubsection{Public data} Public data (used by the Llama model) predominantly contains a near-deduplicated data set of publicly available code (859 GB)~\cite{roziere2023code}. The data set also consists of natural language data sets related to code (78GB). This data set contains many discussions about code and code snippets included in natural language questions or answers.

\subsubsection{First-party data} \label{firstpartydata} For training on our first-party data, we collected data from \Meta's code repositories and notebooks, \ie first-party data, applying several filters \cite{codecompose2023}:
\begin{itemize}
    \item Rather than crawling the entire repository, we used code that is modified through diffs (\Meta's term for pull requests) checked in by developers as a way of staying close to our end application (\ie writing code in a code editor). This way we avoid training on code that may have been added a long time ago but is never modified.
    \item To keep the training data fresh, we only included diffs that are up to 2 years old, and only kept the latest versions of files to avoid bugs that may have been patched.
    \item For each major target language, we exclude code that is not in production or deprecated. After these filters, our first-party training data included in the order of tens of millions of files amounting to a few billion lines of code across more than 10 languages.
\end{itemize}

\subsubsection{SQL artifacts} \label{sqldata} To specialize the model for the SQL domain, we sourced 9 million SQL queries from our internal data warehouse. These queries are fully verified to make sure they pass syntax and semantic checks, and can be executed without producing any run time errors. Additionally, we sourced the schema information of the source tables that are used in these SQL queries. Schema information includes table names, column names and their data types.

\subsection{Data quality improvements}

Dataset quality played a crucial role in shaping the performance and effectiveness of \sa. We deploy a combination of manual curated and automated techniques for improving data quality. The main components of our data quality pipeline include query quality filtering, query diversity filtering and deduplication of similar or repetitive queries. 

For example for the SQL completion learning task, we have leveraged simple heuristics to enhance the quality of our \sa training dataset. The heuristics that we have used include keeping SQL queries with a certain minimum or maximum length, deduping queries and keeping queries that have been run successfully. For instance deduping resulted in filtering of over 5\% of queries (leaving 10 million queries for training after filtering).


Finally, to improve the dataset diversity we have ensured that we have a representative queries from different table schemas and query complexity levels (e.g., easy, medium, hard) defined by the number of SQL components, selections, and conditions. More specifically we defined the complexity levels defined based on the Spider ~\cite{yu2018spider}.
Using the data quality filtering steps above, we ended with about 10M queries for SQL completion task fine-tuning.  


\begin{figure*}
\centering
\includegraphics[width=1.5\imagewidth]{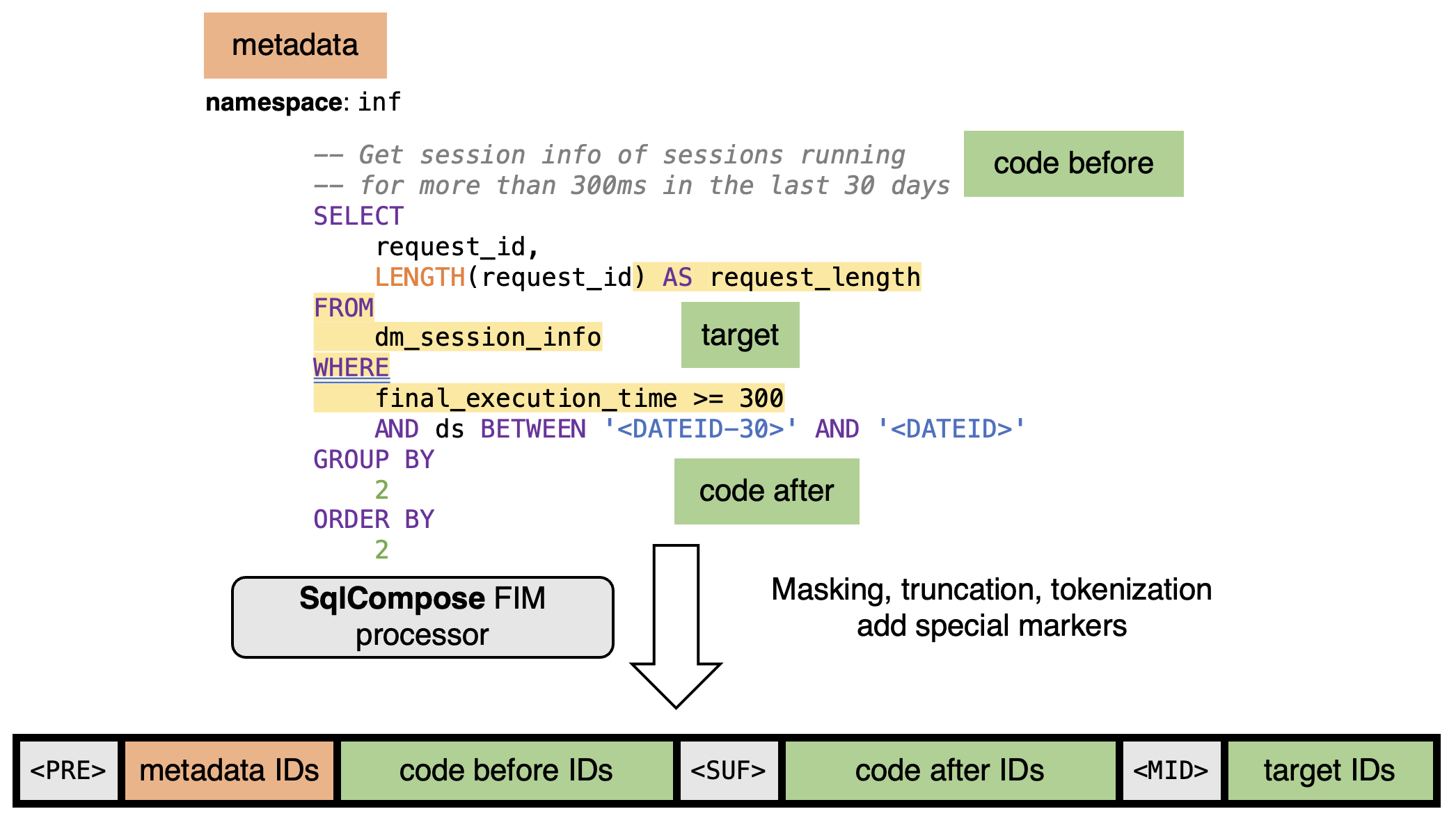}
\caption{Converting a query to fill-in-the-middle (FIM) format with \saFIM's processor}
\label{fig:fim-training}
\end{figure*}

\subsection{\llama model development}
We use a checkpoint of the Llama model that is trained heavily on code as our base model \cite{roziere2023code}. They come in four model sizes: 7B, 13B, 34B and 70B parameters. We take the 7B model as our base model as it helps us strike the right balance between prediction accuracy and end-to-end inference latency requirements of a code completion system ($\sim$200 milliseconds). At the time this work began, we were only able to use the Llama 2 model weights \cite{touvron2023llama} and trained on 500B tokens from a code-heavy data set.

The Llama model is trained predominantly on a near-deduplicated data set of publicly available code. It was trained to support infilling tasks. Infilling is the task of predicting the missing part of a program given a surrounding context. Applications include code completion at the cursor's position in code IDEs, type inference and generation of in-code documentation (e.g., docstrings). Infilling models were trained following the concept of causal masking \cite{aghajanyan2022cm3}, where parts of a training sequence are moved to the end, and the reordered sequence is predicted autoregressively.

\subsection{\saSchema model development}
\saSchema models are trained on top of the \llama models. This model is initialized with Llama model weights \cite{roziere2023code} and trained on first-first party code data. This data includes code from other programming languages such as Python, C++, React, etc. as mentioned in Section \ref{firstpartydata}. This helps the model learn the basics of coding patterns, frameworks, nomenclature used in the company. 

After a checkpoint is produced, we continue to pre-train this checkpoint using the first-party SQL data and SQL schema information to specialize or align the model further on SQL completion and prediction tasks. As the base model has seen SQL queries as part of the pre-training (Llama training), the model already understands how to write SQL queries. However, it does not understand the dialects used at \Meta \ie Presto SQL \cite{presto}. 

Therefore, we continued pre-training on first-party SQL queries and schema. This serves two purposes: it teaches the model about the nuances of the SQL used internally, and it equips the model with the knowledge of the first-party data warehouse, which is very important to prevent the model from hallucinating table and column names while synthesizing SQL queries.

\subsection{\saFIM model development}
\label{subsection_fim_model_dev}
While training the model on the first-party coding data and SQL artifacts, we leverage a training objective named Language Causal Masking (LCM)~\cite{codecompose2023}. This helps the model consume context bidirectionally (code before and code after), which is important for any code completion system. Moreover, LCM overcomes the limitations imposed by regular CM objective with respect to tokenization. We list the modifications we performed to the CM objective to produce LCM objective below:

\begin{itemize}
    \item CM implements the masking after the text has been tokenized into token IDs, which limits the model during training to only seeing mask spans with edges at common tokenizer tokens. LCM lifts the masking step to the language level and avoids this, similar to the fill-in-the-middle (FIM) task~\cite{bavarian2022efficient}. 
    Also, LCM only masks at certain trigger characters -- that is, characters where the model will be queried during inference such as \texttt{(}, \texttt{.}, \texttt{=}, \texttt{SELECT}, \texttt{WHERE}, \texttt{FROM}, etc.
    
    \item We prefix certain metadata to the input in LCM, such as the programming language, full path to the file, and the kernel name for notebooks. 
    
    \item Through model-level ablations, we found an optimal 70-30 split of the model's input length between code before and code after the cursor. 
    
    \item Specialized for our use case, LCM has only one mask in any input. 
\end{itemize}

A step-by-step overview of constructing an input in LCM is shown in Figure~\ref{fig:fim-training}, along with an example SQL query. Once an input is constructed, during training, we maximize the log probability of the language-masked input:
\begin{equation}
\begin{split}
\log \mathcal{P}([\texttt{Metadata}; \texttt{Before}; \texttt{<mask>}; \texttt{After}; \\
\texttt{<mask>}; \texttt{Target}]) 
\end{split}
\end{equation}

where \texttt{Metadata}, \texttt{Before} and \texttt{After} are the tokens in the metadata, code before, and code after the cursor, respectively, \texttt{Target} is the code that was masked, and \texttt{<mask>} is a special token.
During inference, we sample tokens in an auto-regressive manner from the distribution:
\begin{equation}
\mathcal{P}(\cdot~|~[\texttt{Metadata}; \texttt{Before}; \texttt{<mask>}; \texttt{After}; \texttt{<mask>}])
\end{equation}

As we are suggesting lines of code, we stop the generation early once a newline token has been generated. 
Due to the real-time nature of our application and the inline suggestion user experience (UX), we only return one sequence of generated tokens.

\section{Contextual information about \Meta}
\label{sec:BackgroundMeta}
This section offers an overview of the technology stack used by data engineers at \Meta{}. This is important to provide context 
on our dataset and experimental setup. 

\Meta{}'s data warehouse is the main data repository that is used for analytics. It is a collection of millions of tables, 
physically stored using an internal fork of ORC\footnote{Apache ORC, https://orc.apache.org/}
\Meta{}'s exabyte-scale data warehouse is so large that it cannot physically be stored in one single datacenter. Instead, data 
is spread across different geographical locations.

The warehouse, due to its non-centralized nature, is divided into `namespaces.' These namespaces represent both a geographical 
and logical segmentation of the warehouse: tables that share a common ``theme'' are grouped into the same namespace. This allows 
for efficient querying as data does not need to be transferred across different locations. However, if a query requires tables 
from two distinct namespaces (for example, table1 in namespace A and table2 in namespace B), data replication becomes necessary. 
Either table2 can be replicated to namespace A, or table1 to namespace B, allowing the query to be run in the namespace where 
both tables are present. Data engineers have the ability to create these cross-namespace replicas swiftly using a web-based tool, 
and these replicas are automatically synchronized.

Data is typically introduced into the warehouse in three primary ways:
\begin{itemize}
 \item Through data workflows and pipelines, such as data inserted by a Dataswarm pipeline
 This data is usually 
 sourced by querying other tables within the warehouse.
 \item Via logs, which are data produced from either server-side or client-side logging frameworks.
 \item Through daily snapshots of entities present in the production graph database.
\end{itemize}

The warehouse can be queried by many different entry points, but data engineers at \Meta generally use Presto and Spark. While both are
open-source (Presto was originally developed at Meta and was open-sourced in 2019), \Meta{} uses and maintains its own internal forks --- 
but frequently rebases from the open-source repository so that we are kept up-to-date, and contributes features back into the open-source 
projects.

With our focus primarily on business impact, design and optimization, most of our pipelines and queries are written in SQL in one of two 
dialects, Spark SQL or Presto SQL. This approach provides a consistent understanding of the data and business logic and enables any 
data engineer, data scientist, or software engineer comfortable with SQL to understand all of our pipelines and even write their own 
queries. 

The choice of Presto or Spark depends mostly on the workload: Presto is typically more efficient and is used for most queries while Spark 
is employed for heavy workloads that require higher amounts of memory or expensive joins. Presto clusters are sized in a way that most 
day-to-day adhoc queries (that scan, generally, a few billions rows — which is considered a light query at \Meta{} scale) produce results in 
a few seconds (or minutes, if there’s complex joins or aggregations involved).
\subsection{Real-time Querying}
Scuba is \Meta{}’s real-time data analytics framework. It is frequently used by data engineers and software engineers to analyze trends on 
logging data in real time. It is also extensively used for debugging purposes by software and production engineers.

Scuba tables can be queried either through the Scuba web UI (which is comparable to tools like Kibana), or via a dialect of SQL. In the 
Scuba web UI, engineers can quickly visualize trends on a log table without having to write any queries, with data that was generated 
in the past few minutes.
\subsection{Bento and Daiquery}
Daiquery is one of the tools data engineers use on a daily basis at \Meta{}. It is a web-based notebooks experience which 
acts as a single entrypoint to query any data source: the warehouse (either through Presto or Spark), Scuba, and plenty 
of others. It includes a notebook interface with multiple query cells, and users can quickly run and iterate on queries against 
our data warehouse. Results appear as tables by default, but built-in visualization tools allow the creation of many different 
types of plots.

Daiquery is optimized for rapid query development, but does not support more complex post-query analysis. For this, users can 
promote their Daiquery notebooks into Bento notebooks. Bento is \Meta{}’s  implementation of managed Jupyter notebooks, 
and in addition to queries also enables python or R code (with a range of custom kernels for different use cases) and access 
to a wide range of visualization libraries. In addition to its use by data engineers, Bento is also used extensively by data 
scientists for analytics and machine learning engineers for running experiments and managing workflows.

\section{Data, Evaluation Methodology, and Measures}
\label{sec:methodology}

\subsection{Internal Benchmark}
\label{subsection_internal_benchmark}

Benchmarking and evaluation which are essential tools for guiding model improvements through better training data selection, prompt engineering and supervised fine-tuning. Internal benchmarking is especially vital for understanding the performance of the current models, comparing different models (large versus small, catch-all models versus expert models), and even catching future regressions in production deployments of models.
While external data sets such as Spider~\cite{yu2018spider}, Geo Query \cite{Geo} exist for benchmarking general SQL completion, Spider~\cite{yu2018spider} contains annotations from 11 college students. For problems like SQL completion, State-Of-The-Art (SOTA) benchmarking does not even exist, which makes it more difficult to benchmark these applications. Typically, a benchmarking exercise constitutes two primary components: ground truth data, metrics. 

Benchmarking comprises two primary components: ground truth data and metrics.
Ground truth data can be human curated or can be generated programmatically. In both cases, the dataset quality is of utmost importance. To obtain human-generated dataset, we employ human annotators (preferably, domain experts) to curate and annotate datasets with the appropriate labels, responses, etc. While this is of higher quality, it is also resource-intensive. To programmatically obtain ground truth data, we apply heuristics as well as simple machine learning models to curate data sets. Programmatic datasets are of lower quality compared to human-generated datasets, however, scalable and less resource consuming. 

To curate our benchmark, named \SeqComp, we started with a partial held-out dataset from the \saSchema fine tuning data, and randomly cut-off a part of the queries (both consisting of a single line and multi lines) starting from specific trigger characters (e.g., whitespaces, comma, parenthesis). We also categorized each data point by the length of the queries, \ie Small, Medium, and Large, and the complexity of the queries. To classify the queries we defined query complexity levels inspired by Spider ~\cite{yu2018spider} where we classify SQL keywords such as "SELECT", "FROM", and "WHERE" into \textit{easy}, keywords such as "JOIN", "GROUP BY", "HAVING", and "ORDER BY" into \textit{medium}, keywords such as "UNION", "EXCEPT", "INTERSECT", and "LIMIT" into \textit{hard}, and keywords such as "WITH", "CASE", "IF", and "COALESCE" into \textit{extra hard} complexity levels. Regarding the length of the queries, we defined \textit{Small}, \textit{Medium}, and \textit{Large} based on the third quantiles of the dataset.
Then, we randomly selected a subset of the dataset which forms a uniform distribution of all query lengths and complexity levels resulting in a balanced \SeqComp dataset of 15,256 data points. 

{\bf Metrics} help us track the performance of various solutions against the ground truth data. There exist a plethora of offline metrics such as Exact Match (EM), BLEU score, Levenshtein edit distance, ROUGE score that evaluate the performance of translation systems. Further, they can be extended to measure any system that generates text. While the conservative metrics are standard for translation and other text and code generation applications, accomplishing a data task using SQL can be done in an infinite number of ways by querying different tables and columns. Therefore, there is a compelling need for domain-specific metrics such as “containment” or "Table Match" to better understand the usefulness of generated SQL query. To evaluate \saSchema, we created our \SeqComp SQL completion benchmark, and measured a series of standard text and code generation metrics as well as defining SQL-specific metrics.

\subsection{Evaluation Method}
\label{subsection_evaluation_method}
To evaluate \saSchema, \saFIM, and \llama capabilities in predicting SQL completions, we ran them against the \SeqComp benchmark. More specifically, we ask the models to predict the masked part of a given SQL query, such as the ``target'' in Figure~\ref{fig:fim-training}, having the query texts before and after the masked target as their input. We evaluate each model in two modes of single-line, where the model is expected to only complete the query until the end of the first line of the masked target, and multi-line, where the model is expected to complete the whole masked target. We then measure the following metrics:

\begin{itemize}
    \item Exact Match (EM) is a simple but strict metric that evaluates how often the generated SQL is exactly the same as the masked portion.

    
    \item BLEU Score is a ratio measuring the average number of n-grams between the masked portion and the generated SQL. 

    
    \item Containment Score (CS) measures to what extent the predicted completion contains the same SQL keywords as the masked target, such as WHERE clauses, predicates, joins, group, orderby.

    \item Table Match Score (TMS) is a binary score that measures whether the predicted completion contains the same table names as the masked target. This is important to understand how often the models hallucinate table information.
\end{itemize}

\subsection{Evaluation methodology for use in production}

We used a mixed methods approach~\cite{Creswell2017MixedMethods} to evaluate \sa in production, collecting usage data and feedback comments.

{\bf Our rollout strategy} for \sa consists of gradual deployment in waves of randomly selected user cohorts. Within each wave, we rolled it out to increments of 25\% of the developer population until we enable it for 100\% of developers. The rollout was completed after four weeks in the fall of 2023.

We report how many suggestions are accepted by engineers and what proportion of SQL code is written by \sa.
We instrumented telemetry to track various events in the SQL authoring tool such as displaying a suggestion inline, accepting or rejecting a suggestion, and the length of accepted suggestions. In total, our large-scale deployment resulted in \sa making \numSuggestionsActivity suggestions. \numDevsActivity distinct developers have seen at least one \sa suggestion. We only count suggestions that were displayed for at least 750 milliseconds to ensure that developers were exposed to a suggestion and had a chance to see and comprehend it~\cite{codecompose2023}.

Our outcome measures are the acceptance rate of suggestions and the percentage of code typed using \sa. These measures have been used in prior work, with, for example, Google reporting that 3\% of the code typed by engineers was from their AI~\cite{googleblog}.

While we do not use a formal thematic or grounded theory research methodology to understand user feedback, we provide examples of both negative and positive feedback. We have used this feedback to incrementally improve \sa. We also extract overarching themes from the feedback. Future work is necessary to systematically understand the user experience of AI-assisted SQL editing. 

\section{Results}
\label{sec:results}

\begin{table*}
\centering
\caption{Outcome metrics for each model and the percentage point (\pp) change among models. \saSchema performs well on single-line completions. On single-line \saFIM performs similar to \saSchema only making significant improvement on TMS. However, for multi-line predictions, only \saFIM is capable of making high quality suggestions. }
  \begin{tabular}{l|r|rr|rrr}

\toprule


{\bf Single line}	&	\llama	&	\saSchema	&	 vs \llama	&	\saFIM	&	vs \llama	&	vs \saSchema	\\ \hline
EM	&	29	\% &	48	\% &	19	\pp &	50	\% &	21	\pp &	1	\pp \\
BLEU	&	53	\% &	69	\% &	16	\pp &	69	\% &	17	\pp &	1	\pp \\
CS	&	66	\% &	78	\% &	12	\pp &	78	\% &	12	\pp &	0	\pp \\
TMS	&	12	\% &	13	\% &	1	\pp &	23	\% &	11	\pp &	10	\pp \\ \bottomrule \toprule
{\bf Multi-line}	&	\llama	&	\saSchema	&	 vs \llama	&	\saFIM	&	 vs \llama	&	vs \saSchema	\\ \hline
EM	&	0	\% &	0	\% &	0	\pp &	20	\% &	20	\pp &	20	\pp \\
BLEU	&	12	\% &	24	\% &	12	\pp &	59	\% &	47	\pp &	35	\pp \\
CS	&	57	\% &	77	\% &	19	\pp &	82	\% &	25	\pp &	6	\pp \\
TMS	&	26	\% &	62	\% &	36	\pp &	75	\% &	48	\pp &	13	\pp \\

 \bottomrule
  \end{tabular}
  \label{tab:EM-BLEU}
\end{table*}

\subsection{RQ 1. Public \llama}
\textit{How well does the a public model generate SQL code?}

Our goal is to understand how well LLMs work on SQL, which is a declarative language. To establish a baseline model at \Meta, we evaluate how well the public \llama~\cite{roziere2023code} performs at autocompleting SQL queries. 

\llama is an open-sourced model trained on publicly available data. We evaluated it against our internal benchmark, described in Section~\ref{subsection_internal_benchmark}, for both single-line and multi-line SQL completion tasks. For more details on the evaluation methodology and execution refer to Section~\ref{subsection_evaluation_method}.

In Table~\ref{tab:EM-BLEU}, we see \llama's performance in both single-line and multi-line SQL completion tasks. For \llama we see an exact match, BLEU, containment, and table match of 29\%, 53\%, 66\%, and 12\% for single line. The corresponding values for multi-line are 0\%, 12\%, 57\%, and 26\%, respectively. These results are comparable with prior work examining imperative languages~\cite{codecompose2023}. 


As shown in Table~\ref{tab:EM-BLEU}, \llama is able to accurately predict the single-line completion of a small portion of the SQL queries in our internal benchmark. For multi-line, \llama was not able to correctly predict any of the SQL queries in our benchmark, as multi-line completion is a more challenging task which requires accurate table and column names beyond SQL keywords. Additionally, it is not able to predict the table names in many SQL queries of our internal benchmark, \ie only 12\% in single-line and 26\% in multi-line completions.
Note that, single-line completions in our benchmark contain less number of completions with a table name in them. more specifically, since the model is asked to complete only a partial line of a SQL query, that specific line might not be the line containing the table name in the SQL query. Therefore, in multi-line completions the model is presented with more opportunity to predict the table names.

\begin{tcolorbox}
The results for \llama represent a baseline performance for a model as it has not been trained on our internal data, and hence, it is not familiar with the internal SQL queries, coding styles, and table names.
\end{tcolorbox}

\subsection{RQ 2. \saSchema}
\textit{How important is fine-tuning on table schemas?}

To evaluate the impact of fine-tuning in SQL completion task, we fine-tuned \saSchema on first-party data and code at \Meta. LLMs notoriously hallucinate, and in this case, they especially hallucinate table and column names. Therefore, we also fine-tune \saSchema on internal table schema including table and column names.
We evaluate the fine-tuned model, \ie \saSchema, against our internal benchmark, described in Section~\ref{subsection_internal_benchmark}, and measure the same set of metrics as described in Section~\ref{subsection_evaluation_method}.

In Table~\ref{tab:EM-BLEU}, we see the detailed results of \saSchema's performance in single-line and multi-line SQL completion tasks compared to those of \llama, in terms of percentage points (pp) difference. For \saSchema we see an exact match, BLEU, containment, and table match of 48\%, 69\%, 78\%, 13\% for single line. The corresponding values for multi-line are 0\%, 24\%, 77\%, and 62\%, respectively, representing a substantial pp increase over public Llama. 

\begin{tcolorbox}
\saSchema outperforms \llama in both single-line and multi-line SQL completions, except for EM in multi-line which shows that \saSchema is not able to correctly predict any of the multi-line SQL completions either. However, \saSchema's accuracy in predicting the correct table name has significantly improved with a fine-tuning on table schemas, \ie 36 percentage points increased in multi-line. As a result, fine-tuning the model on the first-party data and code significantly improves the performance of SQL completion in our internal benchmark.
\end{tcolorbox}

\subsection{RQ 3. \saFIM}
\textit{How well does a fill-in-the-middle (FIM) model perform?}

SQL authoring is often not linear or sequential. When authoring long queries, it is common for developers to jump around nested sub-queries and common table expressions (CTEs), such that information in the suffix becomes just as important as information in the prefix, \ie they Fill In the Middle (FIM). 

\saFIM is trained on the first-party data and code with FIM objective where the model consumes bidirectional contexts of the code and is asked to fill in the middle. See Section~\ref{subsection_fim_model_dev} for more details of the \saFIM's development. To evaluate the impact of the FIM training we evaluated \saFIM against our internal benchmark and compared its results with \saSchema. 

In Table~\ref{tab:EM-BLEU}, we see the detailed results of \saFIM's performance in SQL completion in both single-line and multi-line modes compared to those of \saSchema in terms of percentage points (pp) difference.
For \saFIM we see an exact match, BLEU, containment, and table match of 50\%, 69\%, 78\%, 23\% for single line. The corresponding values for multi-line are 20\%, 59\%, 82\%, and 75\%, respectively. The improvement in single line is mostly contained to better table match percentages over \saSchema, the multi-line improvement is dramatic, increasing from 0\% exact matches to 20\%. Furthermore, \saFIM suggests the correct table 75\% of the time.

\begin{tcolorbox}
\saFIM outperforms \saSchema in both single-line and multi-line SQL completion tasks. More specifically, we see a significant lift in multi-line completions, \ie an increase of 20 pp in EM from 0\% in \saSchema. Note that EM is the most restrictive metric and it is exceedingly difficult to achieve in multi-line completions, as they include longer responses and more chance of failure. 
\end{tcolorbox}
\begin{tcolorbox}
(continued from above) Additionally, \saFIM is able to predict correct table names in 75\% of the multi-line completions which highlights the necessity of consuming the bidirectional context, \ie the suffix as well as the prefix in a SQL query.
\end{tcolorbox}

\subsection{RQ 4. Adoption and Feedback} \label{adoption}
\textit{How is \sa used in practice?}

\sa has enjoyed a wide and consistent adoption among employees at \Meta. It peaks at approximately 8,100 Daily Active Users (DAU) and 15,700 Weekly Active Users (WAU), where active users are those that accept suggestions consistently. 

\begin{table}
\centering
\caption{\sa Weekly Active Users by Job Family} 
\begin{tabular}{ ll } 
 \hline
 \textbf{Job Family} & \textbf{Weekly Active Users} \\
 \hline
 Software Engineering & 9,120 \\
 
 Data Science & 1,690 \\ 
 
 Data Engineering & 1,260 \\ 
 
 Production Engineering & 560 \\ 
 
 Other & 3,060 \\ 
 \hline
\end{tabular}
\end{table}

In the first quarter of 2024, the system made over 8 million suggestions at a rate of approximately 680,000 suggestions per week. On average, users accepted 21\% of the suggestions that were shown for more than 750 milliseconds.
This makes up over 50 million characters of SQL and represents almost 6\% of all the SQL authored at \Meta.

While acceptance rate helps us understand the likeability of \sa and can serve as a proxy to suggestion accuracy, it can be gamed easily. 
We propose a new metric named Characters accepted Per Opportunity (CPO) \cite{cpo}. An opportunity is any editing action in the editor that could trigger a suggestion. 
CPO allows us to track the throughput of accepted suggestions in a normalized way. It is more robust than simple acceptance rate because it also takes into account the length of accepted suggestions and cannot be affected by simply showing less and/or trivial suggestions. The system records a CPO of 2.2. 

Week-over-week product retention, which is the fraction of weekly active users who were also active in the previous week, for \sa is 80\%. Opt-out rate, which is the number of users that disabled \sa voluntarily. The opt-out rate of \sa currently ranges around 0.3\%.

\subsection{User sentiment and feedback}
At \Meta, developers are encouraged to post their feedback in a tool-specific feedback group. The developers are generally vocal and provide feedback despite the feedback being publicly visible to others, including the developers of \sa. We use the user feedback to keep track of sentiment, learn about UX and suggestion accuracy issues, and identify bugs.

\subsubsection{Favorable scenarios for \sa}
The scenarios for which \sa was able to add the biggest value includes (but is not limited to), completing tedious or repetitive SQL clauses, boilerplate coding, helping eliminate the need to remember difficult SQL syntax. We also noticed auxiliary benefits reported by the developers that \sa helped them in filling out the natural language parts of the query (e.g., column alias names or inline comments). 

Many developers highlighted the fact that the suggestions are accurate. Also, we received feedback about how nicely \sa navigates the precision versus recall problem by not showing suggestions too often, which is reflected in the metrics we shared in Section \ref{adoption}. Also, we noticed that \sa is being received by occasional and experienced SQL developers alike. We list some anecdotes below.

\begin{quote}
\textit{``The other day it literally wrote the exact query I wanted, which was one with a window function which I always forget the syntax for. For me as a PM, who only digs into data every now and then, it really makes it more efficient as I don't have to pull up the Presto SQL documentation to see how functions are working all the time."}
\end{quote}

This summarizes the productive experience a product manager who does not author SQL frequently.

\begin{quote}
\textit{``I just wanted to give a quick shout out that inline completions has been getting really good and I've personally seen the improvements as someone who writes a lot of queries. Today, I was writing a comment, and I literally only wrote one word and it somehow telepathically knew exactly what I wanted to say, and auto complete was great. Also, queries auto completion has been getting noticeably better with picking up patterns, assigning date stamps, writing filters, etc."}
\end{quote}

Here, the feedback summarizes the delightful and productive experience a data engineer who authors SQL regularly had with \sa.

Another fascinating trend we noticed in these feedback items is, developers tend to be fine with reworking the suggested queries per their needs as long as \sa helps them get started. As SQL development is iterative, the biggest time savings and productivity boost comes from the fact that \sa helps developers bootstrap and iterate on complex SQL queries without needing to navigate clunky SQL syntax, semantics, and documentation.

\begin{quote}
    \textit{``\sa saved me hours of SQL iteration today: I feel like in its current state, the value of \sa to me is already there. The answers in here aren't exactly right, but they are close enough to have saved me a lot of smashing my face into daiquery which is what usually happens when I start doing some new data analysis. \sa instantly got me from 0 to 10 to about 90\%, and I was able to do the rest on my own."}
\end{quote}

Additionally, many developers provided unsolicited feedback about how helpful \sa has been with respect to saving several hours of their time and contributing towards bringing a positive developer experience at \Meta. 

\begin{quote}
\textit{``I just wanted to drop a line and say, I got to use this feature today. The auto-complete and suggestions are a fantastic use of generative AI."}
\end{quote}

\begin{quote}
    \textit{``This query would have taken me several hours, but with \sa, it saved me a significant amount of time. This is awesome!"}
\end{quote}

\subsubsection{Unfavorable scenarios}
After analyzing negative feedback, we found that common problems themed around hallucinating table names, issues related to UX such as \sa competing with traditional inline completions, etc. 

A developer passed the following feedback about \sa:
\begin{quote}
    \textit{``I was playing around to see if it could write queries for me in SQL and it looks like it is hallucinating columns that don't exists on the tables."}
\end{quote}

\begin{quote}
    \textit{``I attempted to use \sa to help with writing a daiquery query since I am bad at SQL, but it seems to have no knowledge of any of the tables"}
\end{quote}

While we reduced hallucinations significantly (as reported in Table \ref{tab:EM-BLEU}), it is hard to solve it completely as tables keep getting moved, renamed, deleted, and created continuously. The scale at which these operations happen at \Meta amplifies the complexity further. Our offline tests shows that table names are correct around 75\% of the time. 

Additionally, SQL is a language where \texttt{SELECT} (and the column list) is authored before writing the table names many times. This makes the problem even more difficult. To alleviate that, we train the models on schema information and the model tend to perform significantly better in its ability to pass subsequent suggestions once the developer writes the \texttt{FROM} clause and the table names.

Another developer expressed their negative experience about overloading of the tab key for both indentation and accepting suggestions:
\begin{quote}
\textit{"I have a distinct style of formatting my SQL queries that I've been using for 5+ years and will likely never change, as it makes SQL much more readable for me. As part of this I utilize the tab key extensively for indenting + spacing. As you can probably imagine, this makes tab-autocomplete a frustrating experience for me, especially when it triggers really quickly."}
\end{quote}

Coexisting with the traditional auto complete system is a challenge faced by many AI-assisted code authoring solutions~\cite{codecompose2023}. Developers tend to have strong preferences around UX, keyboard shortcuts, and the stylistic aspects. It takes time, great amount of user education, and novel and innovative ways if presenting AI suggestions to make the experience enjoyable for everyone.


\begin{tcolorbox}

\sa has is used on a weekly basis by over 10k users including data scientists and software engineers, less than 1\% of users have disabled \sa. We use the feedback from users to improve \sa. Interesting positive themes include completing tedious or repetitive SQL clauses, suggesting boilerplate
coding, and help in eliminate the need to remember difficult SQL syntax. The most significant negative themes was table and column name hallucinations, which has been reduced with the release of \saFIM. Other negative themes include interfering with traditional auto-complete system and changes in the keyboard shortcuts and the stylistic aspects.  

\end{tcolorbox}

\section{Threats to Validity}
\label{sec:threats}

\subsection{Generalizability}

Drawing general conclusions from empirical studies in software engineering is difficult because any process depends on a potentially large number of relevant context variables. The analyses in the present paper were performed at \Meta, and it is possible that results might not hold true elsewhere. 
However, our study does cover a very wide swath of software engineering. The software systems covers millions of lines of code and 10's of thousands of developers who are both collocated and working at multiple locations across the world.
We also cover a wide range of domains from user facing social network products and virtual and augmented reality projects to software engineering infrastructure, such as calendar, task, and release engineering tooling. 

In Sections~\ref{sec:model} and \ref{sec:methodology} we provide detailed steps and descriptions of the types of data we used in our model. We look forward to reading how other researchers use LLMs to assist in SQL composition. 

\subsection{Construct Validity}

To evaluate our models in offline tests, we used standard metrics such as exact match and BLEU scores. However, given that SQL is declarative and is often not written in a sequential manner, we introduced two new metrics (see Section~\ref{sec:methodology}). The Containment Score (CM), which determines how many SQL clauses are correct. We also introduced the Table Match Score. A hallucinated table name will drastically reduce the quality of the SQL and impact column names. These metrics need further validation and we hope that other researchers will build upon our SQL specific measures. 

\subsection{Internal Validity}

Unlike a traditional experiment, we also have to produce and release a running product. While our offline experimental results and benchmark are on consistent dataset, our rollout is an ongoing process without a constrained timeframe. Instead of a traditional research method, we monitor the feedback and usage results on a daily basis. New features are gradually rolled-out to avoid any regressions using an A/B test methodology~\cite{NudgeBot2022}. 

\section{Literature and Discussion}
\label{sec:literatureAndDiscussion}
To the best of our knowledge, almost all published work in the domain of authoring SQL using LLMs is focused on the problem of converting natural language to SQL queries (Text2SQL)~\cite{zhang2020m,li2024pet,zhang2024benchmarking,zhong2017seq2sql}, whereas the focus of this work is on autocompleting SQL queries as authors type them out. While related, the problems are different and need different approaches. For instance, (a) for autocompleting SQL queries, the LLM must be trained on a large number of SQL queries instead of text-sql pairs, (b) the latency constraints are much tighter for an autocompletion tool compared to natural language querying, and (c) the model must be able to pick up context from after the cursor unlike the relatively straight-forward left-to-right generation in the typical Text2SQL setting.

\textbf{Text2SQL:} Early works did not handle generalization to unseen databases well, later works such as RAT-SQL did attempt to generalize to unseen database schemas but assumed the schemas are small enough that they can be encoded at query time, and furthermore assumed that the system knows which database to look at~\cite{wang2019rat}.





Most works assume small database schema or the schema is known at runtime, which is not the case for us~\cite{wang2019rat}. text2sql assumes the model is aware of the intent of the author, whereas in completion even the intent is not clear in most of the cases. makes things like encoding schema in the query, or using PICARD (constrained decoding) hard. Even if these were known in a subset of cases, latency constraints make it hard to use them. Some works also use the DB content which cannot work for us due to the sheer number of dbs/tables and size of the data, and the latency constraints. Further complication in our setting is there may be multiple tables that may answer the user's question which makes evaluation also challenging.

\textbf{Code completion:} While there is a large body of work around code autocompletion with LLMs\cite{bruch2009learning,robles2008program,proksch2015intelligent,zhou22improving,kim21code}, there has been limited deployment of these in large industrial environments \cite{intellicode,codewhisperer,copilot,googleblog,codecomposemultiline}. These works have been effective at generating code in programming languages. For example, Nguyen et al. \cite{nguyen2022empirical} used 33 LeetCode questions to create queries for Copilot in four different programming languages. They found that Copilot's Java suggestions have the highest correctness score (57\%) while JavaScript is the lowest (27\%).

Although there are blog posts explaining to developers how to use GitHub Copilot with SQL~\cite{CopilotForSQL2022,CopilotAzureData2023}, there is no description of the model and no evaluation of how well it performs on SQL. Our paper illustrates three main challenges that warrant special treatment for SQL autocompletion: (i) its declarative nature coupled with its ties to a data warehouse, (ii) exacerbated impact of LLM hallucinations, and (iii) atypical coding styles (CTEs, non-linear authoring) when developers write SQL queries. In this light, our work is the first to develop specialized models for AI-assisted SQL authoring at scale with comprehensive evaluation.

\section{Conclusion and Contributions}
\label{sec:conclusion}

Our major contribution is to show how well LLMs can work in the context of SQL. Our specific contributions that provide answer to our research questions are the following:

\begin{enumerate}

\item RQ 1. Public \llama: We see an exact match, BLEU, containment, and table match of 29\%, 53\%, 66\%, and 12\% for single line. The corresponding values for multi-line are 0\%, 12\%, 57\%, and 26\%, respectively. These results are comparable with prior work examining imperative languages like python~\cite{codecompose2023}. 

\item RQ 2. \saSchema: We see an exact match, BLEU, containment, and table match of 48\%, 69\%, 78\%, 13\% for single line. The corresponding values for multi-line are 0\%, 24\%, 77\%, and 62\%, respectively. These results represent a substantial improvement over the public \llama model. 

\item RQ 3. \saFIM: We see an exact match, BLEU, containment, and table match of 50\%, 69\%, 78\%, 23\% for single line. The corresponding values for multi-line are 20\%, 59\%, 82\%, and 75\%, respectively. The improvement in single line is mostly contained to better table match percentages over \saSchema, the multi-line improvement is dramatic, increasing from 0\% exact matches to 20\%. Furthermore, \saFIM suggests the correct table 75\% of the time. 

\item RQ 4. Rollout and Feedback: \sa has is used on a weekly basis by over 10k users including data scientists and software engineers, less than 1\% of users have disabled \sa. We use the feedback from users to improve \sa. Interesting positive themes include completing tedious or repetitive SQL clauses, suggesting boilerplate
coding, and help in eliminate the need to remember difficult SQL syntax. The most significant negative themes was table and column name hallucinations, which has been reduced with the release of \saFIM. Other negative themes include interfering with traditional auto-complete system and changes in the keyboard shortcuts and the stylistic aspects.  

\end{enumerate}

\balance
We anticipate that other researchers will build upon our techniques, models, and evaluation metrics to ensure LLMs continue to accelerate and assist in writing SQL.

\section*{Acknowledgements}
We would like to thank Charlie Regan, Kristian Kristensen, Daniel Cheng, Kelly Hirano, Bhaskar Mehta, Barak Yagour, Killian Murphy, Aparna Ramani, Dinkar Pataballa, Shahin Sefati, Michael Jiang, Emily Yeh, Kamran Asif, Peyton Foucht, Mariel Carter, Imad Ahmad, Gabriel Synnaeve, Baptiste Rozière, Jeremy Reizenstein, Sten Sootla, Maria Lomeli, and Michael Bolin for their help and support with this work. 

\bibliographystyle{IEEEtran}
\bibliography{ref.bib}

\end{document}

%% file: PeterMacros.tex
\usepackage{xcolor}
\usepackage{xspace}

\newcommand{\pp}{pp\xspace}

\newcommand{\ie}{\hbox{\emph{i.e.}}\xspace}

\PackageWarning{TODO:}{X}
\PackageWarning{TODO:}{X\%}

\usepackage{tcolorbox}

%% file: Macros.tex
\usepackage{xspace}
\usepackage{dirtytalk}
\usepackage{xcolor}

\newcommand{\Meta}{Meta\xspace}

\newcommand{\sa}{SqlCompose\xspace}
\newcommand{\saSchema}{SqlComposeSA\xspace}
\newcommand{\saFIM}{SqlComposeFIM\xspace}
\newcommand{\llama}{Llama\xspace}

\newcommand{\SeqComp}{SeqComp\xspace}

\usepackage{xspace}
\usepackage{dirtytalk}
\usepackage{xcolor}

\definecolor{teal}{RGB}{0,128,128}

\newcommand{\DefMacro}[2]{\expandafter\newcommand\csname rmk-#1\endcsname{#2}}
\newcommand{\UseMacro}[1]{\csname rmk-#1\endcsname}
\newcommand{\rom}[1]{\uppercase\expandafter{\romannumeral #1\relax}}

\newcommand{\numDevsActivity}{16 thousand\xspace}
\newcommand{\numSuggestionsActivity}{4.5 million\xspace}